\documentclass[a4paper,12pt]{article}
\usepackage{amsfonts,amsmath,amssymb}
\usepackage{graphicx,color}
\usepackage{epsfig}

\begin{document}

\title{Four-point correlator of vector currents and electric current susceptibility in holographic QCD}

\author{A. Krikun (ITEP,MIPT)  }


\date{\today}

\maketitle

\begin{abstract}
In this work we calculate the four-point correlation function of vector quark currents of QCD via holographic QCD model. Computing the correlator we take into account the exchange of vector and axial vector  bosons and dilaton in the bulk. The result is used for calculation of the two-point correlator of electromagnetic currents in external magnetic field at zero momentum, related to RHIC experiments, chiral magnetic effect and lattice study. At zero temperature we find this quantity to be loosely connected with chiral symmetry breaking and strongly dependent on the confinement properties. Some features of the AdS/QCD models are also discussed.
\end{abstract}

\newcommand{\p}{\partial}
\newcommand{\M}{\mathcal{M} \left(-\frac{1}{4} q^2, \frac{1}{2}, t^2 \right)}
\newcommand{\W}{\mathcal{W} \left(-\frac{1}{4} q^2, \frac{1}{2}, t^2 \right)}
\newcommand{\ph}{\varphi}
\newcommand{\la}{\langle}
\newcommand{\ra}{\rangle}
\newcommand{\lr}{\leftrightarrow}
\newcommand{\lar}{\leftarrow}
\newcommand{\rar}{\rightarrow}
\newcommand{\paral}{\parallel}

\section{Introduction}
Nowadays the holographic approach to QCD is a common tool for exploring the strong interaction properties at nonperturbative region. There is a number of models \cite{Erlich,Pomarol,soft,Gherghetta,Andreev1}, which can reproduce meson decay constants, Regge trajectories of meson masses, phenomena of spontaneous symmetry breaking and chiral anomaly, confinement of quarks etc. At the same time the model which would incorporate all these features simultaneously, does not exist. Although holographic models can not yet provide a precise calculation tool, they can give some new hints about the nature of the considered phenomena and reveal the connection between different features of the theory. Moreover, the features of the model should be studied to clarify the further modifications.

The paper concerns the calculation of the 4-point correlator of vector currents of QCD in the framework of the holographic model containing the vector, axial vector, scalar and dilaton dynamical fields. Similar strong coupling calculations were made in different models, where the photon correlators in supersymmetric Yang-Mills or conformal QCD \cite{Hatta,Yoshida,Cornalba} and various features  of low energy QCD \cite{Hassanain1, Hassanain2, Konush} were studied. We assume that scalar and dilaton field have some arbitrary background profiles, moreover we incorporate the arbitrary deformation function of the metric in the infrared region. Thus we get the result, applicable to a large class of particular holographic models which appear in the literature. For the perturbative calculation we adopt a convenient Witten diagram technique to represent the action on the classical solutions as the sum of tree diagrams.

The second goal of this work is to study the electric current magnetic susceptibility, namely the electric current two-point function in the external magnetic field at vanishing momentum. This entity may be related to the measured charge asymmetry in heavy ion collisions at RHIC experiment \cite{Kharzeev}, and is studied extensively. According to \cite{Kharzeev}, the susceptibility is generated in the moment of collision by the chiral magnetic effect and the chiral chemical potential. The other input is produced by   the nonzero temperature of the fireball. We study this object at zero temperature and zero chiral chemical potential. Though this setting is not sufficient to study the phenomena in the ion collision, we try to find out if any other origins of current susceptibility are present in nonperturbative QCD. We see that the susceptibility in the holographic model is nonzero at quadratic order in the field, and it is directly related to the 4-point correlator computed in the first part.

As the main result we make some conclusions about the behaviour of the quantity of interest. It turns out that the susceptibility at zero temperature behaves as $Q^4$ at small $Q$. It is not very sensitive to the pattern of chiral symmetry breaking present in the model, namely it is nonzero even when the quark mass and condensate are taken as zero, but depends strongly on the way the conformal symmetry is violated, or of the confinement properties. 

As to the perturbation theory in the bulk space of AdS/QCD, we find that usually the exchange of the dilaton makes no difference being suppressed by the factor of $N_c$. Also we find that the exponential growth of the bulk-to-bulk propagators in the soft wall model with positive dilaton makes the perturbation theory there inconsistent.

In Section 2 we present the setting we are working in, describe the principles of the diagram technique and calculate the four-point function of the electromagnetic currents. In Section 3 we use this result to compute the electric current susceptibility in various models and investigate its dependence on the parameters. The last section is devoted to the discussion of the results obtained.

\section{The vector current four-point function}
\subsection{The setting}
The AdS/QCD model used in this work contains vector ($V^a_\mu$) and axial vector ($A^a_\mu$) fields in the adjoint representation of the gauge group, which is dual to vector currents of QCD ($ \la \bar{q} t^a \gamma_\mu q \ra $ and $ \la \bar{q} t^a \gamma_5 \gamma_\mu q \ra $, respectively), the bifundamental scalar field ($X^{\alpha \beta}$) representing the scalar quark current $\la \bar{q}^\alpha q^\beta \ra$, and a dilaton, which corresponds to the trace of the gluon field strength tensor squared ( $Tr(G^2)$ ). The gauge group is the flavor group $U(N_f)$ of QCD and the model is non-Abelian. Generators of $U(N_f)$ are normalized the standard way:
\begin{equation*}
[t^a,t^b]= i f^{abc} t^c, \qquad \la t^a t^b \ra = \frac{1}{2} \delta^{ab}.
\end{equation*} 
The model is built in 5-dimensional space, which has an AdS metric with equal to unity curvature radius and some arbitrary warp factor $W(z)$. It can be cut at some infrared brane, located at $z=z_m$. $W(z)$ tends to zero at the boundary, in order the space to be the asymptotically AdS.
$$
d^2 s= \frac{e^{-2W(z)}}{z^2} (-dz^2 + dx^{\mu} dx_{\mu}), \qquad z\leq z_m.
$$
The action is a sum of Yang-Mills and Chern-Simons parts. We study the dynamics of the fluctuations of scalar field around its vacuum profile $X_0(z)=\frac{1}{2} \chi(z)$, represented by the pseudoscalar field $\pi$, $X = X_0 e^{i\pi}$. The fact that only axial vector field interacts with this background is a sign of the chiral symmetry breaking in the model. The dilaton field can have a vacuum profile as well. We will denote it by $\Phi(z)$ and consider the fluctuations $\phi(Q,z)$. In the following it is convenient to introduce the sum of the warp factor and dilaton vacuum profile: $\tilde{\Phi}(z)=W(z) + \Phi(z)$. The pure Yang-Mills part is symmetrical with respect to the exchange of left and right currents, the CS part is anti symmetrical.
\begin{align}
\label{S1}
S = \int d^4 x dz  &\sqrt{g} e^{-(\Phi+\phi)} \left( -\frac{N_c}{24 \pi^2} \right) \la F_V^2 + F_A^2 + 6 \chi(z)^2 (A-\p \pi)^2 + \mathcal{A}^2 (\p \phi)^2 +  M(\phi) \ra \\
\notag
+ &\left( \frac{N_c}{12 \pi^2}\right) \la V {\wedge} F_V {\wedge} F_A + V {\wedge} F_A {\wedge} F_V + A {\wedge} F_V {\wedge} F_V \ra
\end{align}
Here the trace is taken over the gauge group $U(N_f)$, $\mathcal{A}$ is a normalization constant of dilaton field, which should be fixed. $M(\phi)$ is a possible potential for dilaton, which is not essential in our calculation, as long as we are not interested in the gluon condensate corrections. All the other constants in the action are fixed by matching with QCD sum rules (see \cite{Erlich,Pomarol,Krik1,Krik2}). In the vicinity of the boundary of the AdS the asymptotics of the scalar vacuum profile is $\chi(z) = mz + \sigma z^3$. The normalization is chosen so that $m$ is the quark mass in QCD.  In the action we have omitted all terms with interactions unnecessary for our calculation. These are interactions of the second order in axial field in the Yang-Mills part and terms of fourth and higher order in fields in the Chern-Simons part.

Before we proceed, we have to fix the constant $\mathcal{A}$ in the action. This can be done by matching with QCD sum rule result for the gluon correlator \cite{Kataev}
\begin{align}
\label{(GG)}
\int d^4 x \la G^2(0)G^2(x) \ra e^{iQx} = -\frac{N_c^2-1}{4 \pi^2} Q^4 ln \left(\frac{Q^2}{\mu^2} \right).
\end{align}
In order to compute this correlator in the holographic model it is enough to solve the linearized equation of motion for the dilaton near the boundary of AdS
\begin{align*}
\p^2_z \phi - \frac{3}{z} \p_z \phi +Q^2 \phi = 0. 
\end{align*}
The solution (bulk-to-boundary propagator) is
\begin{align*}
\phi^{(1)}(z) = \frac{Q^2}{2} [\hat{\phi} z^2 K_2(Qz) + C_2 z^2 I_2(Qz)],
\end{align*}
where $I$ and $K$ are the modified Bessel functions, and $\hat{\phi}$ is a boundary value of dilaton field, which should be associated with the source of corresponding operator in the generating functional of 4D theory, namely $Tr( G^2)$. The final step is to insert this solution into the action and take the second variation with respect to the source $\hat{\phi}(Q)$. We get the two-point function at large momenta $Q$:
\begin{align*}
\frac{\p S}{\p \hat{\phi}(Q) \p \hat{\phi}(-Q)} = - \mathcal{A}^2 \left( \frac{N_c}{24 \pi^2} \right) \frac{1}{8} Q^4 ln(Q^2 \epsilon^2)
\end{align*}
Comparing this with sum rule (\ref{(GG)}) we see:
\begin{align}
\label{A}
\mathcal{A}^2 \ &= \frac{48(N_c^2-1)}{N_c} 
\end{align}

\subsection{The variation of classical action}
We proceed to our main task, the calculation of four-point function of vector currents. According with the general AdS/CFT recipe \cite{Gubser} we take the variation of classical action with respect to four boundary values of vector fields 
\begin{equation*}
\la J_\alpha^a(k_1) J_\beta^b(k_2) J_\gamma^c(k_3) J_\delta^d(k_4) \ra = \frac{\delta^4}{\delta \hat{V}_\alpha^a(k_1) \delta \hat{V}_\beta^b(k_2)\delta \hat{V}_\gamma^c(k_3)\delta \hat{V}_\delta^d(k_4)}S_{cl}(\hat{V}^4).
\end{equation*}

So we need to calculate the classical action up to the fourth order in boundary values, which involves the calculation of classical solutions up to the third order in sources. The calculation can be done perturbatively. Let's make a Fourier transformation over four coordinates and rewrite the action (\ref{S1}) in a more explicit form:
\begin{align*}
S=S^V  + S^\phi + S^A 
\end{align*}
\begin{align}
\left( \frac{24 \pi^2}{N_c} \right) S^V=
\label{S_V-bound}
& \int d^4 q_1 d^4 q_2 \ \delta^4(q_1{+} q_2) \ \frac{e^{-\tilde{\Phi}}}{z} (-) V_\mu^{a}(q_1,z) \p_z V_\mu^{a}(q_2,z) |_{z=0} \\
\label{S_V-kin}
+ &\int d^4 q_1 d^4 q_2 dz \ \delta^4(q_1{+} q_2)  \  V^{a}_\alpha(q_1,z) \frac{e^{-\tilde{\Phi}}}{z} \bigg[-\frac{z}{e^{-\tilde{\Phi}}}\p_z \frac{e^{-\tilde{\Phi}}}{z} \p_z g_{\alpha \mu} -  q_2^2 g_{\alpha \mu} + q_{2\alpha} q_{2\mu} \bigg] V^{a}_\mu (q_2,z) \\ 
\label{S_V-triple}
+ &\int d^4 q_1 d^4 q_2 d^4 q_3 dz \ \frac{e^{-\tilde{\Phi}}}{z}  (2 i) \mathbb{T}^{abc}_{\alpha \beta \gamma} V_\alpha^{a} (q_1,z) V_\beta^{b} (q_2,z) V_\gamma^{c} (q_3,z) \\
\label{S_V-quad}
+ &\int d^4 q_1 d^4 q_2 d^4 q_3 d^4 q_4 dz \ \frac{e^{-\tilde{\Phi}}}{z} \ (-) \frac{1}{2} \mathbb{Q}^{abcd}_{\alpha \beta \gamma \delta} \ V_\alpha^{a} (q_1,z) V_\beta^{b} (q_2,z) V_\gamma^{c} (q_3,z) V_\delta^{d} (q_4,z)  
\end{align}
In this term the partial integration has been used in the kinetic term, and we use the notation for the triple and quad vertex functionals:
\begin{align}
\mathbb{T}_{\alpha \beta \gamma}^{abc} &=\delta^4(q_1 {+} q_2 {+} q_3) \  f^{abc} (g_{\alpha \beta} q_{1\gamma} + g_{\beta \gamma} q_{2\alpha} + g_{\alpha \gamma} q_{3\beta}),\\
\mathbb{Q}_{\alpha \beta \gamma \delta}^{abcd} &=\delta^4(q_1 {+} q_2 {+} q_3 {+} q_4) \  f^{abe} f^{ecd} g_{\alpha \gamma} g_{\beta \delta},
\end{align}
Similarly, the partial integration is made in the dilaton part of the action, but due to the absence of the dilaton source in the problem, the boundary term vanishes, and only kinetic and interaction terms survive
\begin{align}
\label{S_phi-kin}
\left( \frac{24 \pi^2}{N_c} \right) S^\phi = &\int d^4 q_1 d^4 q_2 dz  \ \frac{48(N_c^2-1)}{N_c} \ \phi(q_1,z)  \frac{e^{-\tilde{\Phi}-2W}}{z^3} \bigg[ - \frac{z^3}{e^{-\tilde{\Phi}-2W}}\p_z \frac{e^{-\tilde{\Phi}-2W}}{z^3} \p_z - q_2^2 \bigg] \phi(q_2,z) \\
\label{S_phi-triple}
+ &\int d^4 q_1 d^4 q_2 d^4 q_3 dz \ \frac{e^{-\tilde{\Phi}}}{z} \ \mathbb{D}^{ab}_{\alpha\beta} \ \phi(q_3,z)   V_\alpha^{a} (q_1,z) V_\beta^{b} (q_2,z).
\end{align}
The notation for dilaton vertex functional is
\begin{equation}
\mathbb{D}_{\alpha \beta}^{ab} = \delta^4(q_1 {+} q_2 {+} q_3) \ \Big\{ - \delta^{ab} \p_{z}^1 \p_{z}^2 g_{\alpha \beta} + \delta^{ab} (-(q_1 q_2) g_{\alpha \beta} + q_{1\beta} q_{2\alpha} ) \Big\},
\end{equation}
where by definition momenta $(q_1,q_2,q_3,q_4)$ and differentials $\p_{z}^1, \p_{z}^2,\dots$ correspond to fields with flavor indices $(a,b,c,d)$ respectively.

The $CS$ term warrants more careful study. Its explicit form in the coordinate space is
\begin{align*}
S^{CS} = \int d^4 x dz \frac{N_c}{12 \pi^2} \ 4 \langle t^a t^b t^c \rangle \epsilon^{\mu \nu \rho \sigma}&
(\p_z A_\mu^c [\p_\sigma V_\nu^a V_\rho^b - V_\nu^a \p_\sigma V_\rho^b] \\
&  + A_\mu^c [\p_z V_\nu^a \p_\sigma V_\rho^b - \p_\sigma V_\nu^a \p_z V_\rho^b] \\
& + \p_\sigma A_\mu^c [V_\nu^a \p_z V_\rho^b - \p_z V_\nu^a V_\rho^b]).
\end{align*}
We perform the integration by parts in z coordinate and we omit the boundary value, because the axial source is not present. So in the momentum space the part $S^A $ looks like
\begin{align}
\notag
\left( \frac{24 \pi^2}{N_c} \right) S^A = &\int d^4 q_1 d^4 q_2 dz \ \delta^4(q_1{+} q_2) \times \\
\label{S_A-kin}
  & \times \Big\{ A^{a}_\alpha(q_1,z)  \bigg[-\p_z \frac{e^{-\tilde{\Phi}}}{z} \p_z g_{\alpha \mu} - \frac{e^{-\tilde{\Phi}}}{z}( q_2^2 g_{\alpha \mu} - q_{2\alpha} q_{2\mu}) \bigg] A^{a}_\mu (q_2,z) \\
\notag
 &+ 3 \frac{e^{-\tilde{\Phi}-2W}}{z^3} \chi(z)^2 \Big[ \big(A^{a}_\mu(q_1,z) - q_{1\mu} \pi(q_1,z) \big) \big(A^{a}_\mu(q_2,z) - q_{2\mu} \pi(q_2,z) \big) 
+(\p_z \pi(q,z))^2 \Big] \Big\} \\
\label{S_A-triple}
+ &\int d^4 q_1 d^4 q_2 d^4 q_3 dz \ (24 i )\ \mathbb{A}_{\alpha \beta \gamma}^{abc} V_\alpha^a(z_1,q_1) V_\beta^b(z_2,q_2) A_\gamma^c(z_3,q_3),
\end{align}
where we've introduced the notation for the axial vertex functional
\begin{equation}
 \mathbb{A}_{\alpha \beta \gamma}^{abc} =\delta^4(q_1 {+} q_2 {+} q_3) \ \langle t^a t^b t^c \rangle \epsilon^{\alpha \beta \gamma \sigma}
(\p_z^1 q_{\sigma}^2 - \p_z^2 q_\sigma^1 ).
\end{equation}

\begin{figure}[h]
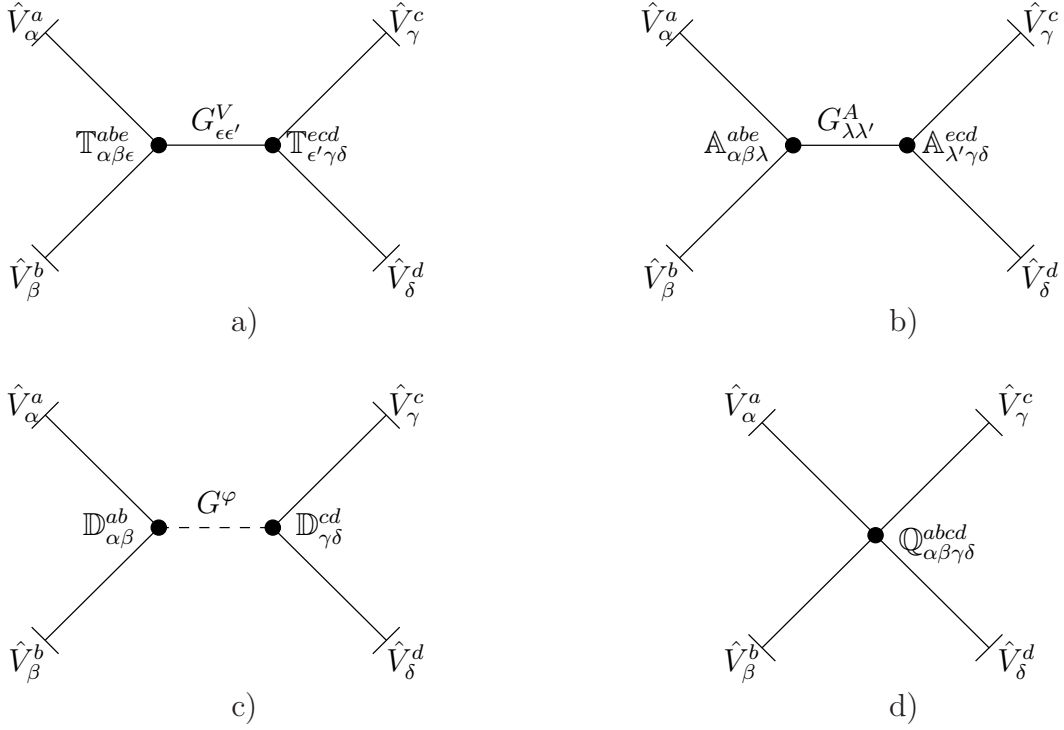

\begin{minipage}[h]{0.5\linewidth}
\center{\input{diagram1a.pstex_t}}
\end{minipage}
\begin{minipage}[h]{0.5\linewidth}
\center{\input{diagram1b.pstex_t}}
\end{minipage}
\begin{minipage}[h]{1\linewidth}
\begin{tabular}{p{0.5\linewidth}p{0.5\linewidth}}
\centering a) & \centering b)\\
\end{tabular}
\end{minipage}
\begin{minipage}[h]{0.5\linewidth}
\center{\input{diagram1c.pstex_t}}
\end{minipage}
\begin{minipage}[h]{0.5\linewidth}
\center{\input{diagram1d.pstex_t}}
\end{minipage}
\begin{minipage}[h]{1\linewidth}
\center
\begin{tabular}{p{0.5\linewidth}p{0.5\linewidth}}
\centering c) & \centering d)\\
\end{tabular}
\end{minipage}
\caption{Tree diagrams, corresponding to the classical action of the fourth order in vector sources: a) vector intermediate boson, b) axial intermediate boson, c) exchange of dilaton, d) quad vector vertex}
\end{figure}

Now we should insert classical solutions calculated order-by-order in vector field sources: $V=V^{(1)} + V^{(2)} + V^{(3)}; A=A^{(2)}; \phi = \phi^{(2)}$. One can check, that the term with $V^{(1)}V^{(3)}$ in (\ref{S_V-kin}) cancels the boundary term (\ref{S_V-bound}). So we are left with the four different diagrams of Fig.1. (1a) comes from the connection of two $V^{(2)}$ via kinetic term in (\ref{S_V-kin}) and connection of two $V^{(1)}$ and one $V^{(2)}$ by the triple vertex in (\ref{S_V-triple}). The connection of two $A^{(2)}$ in the quadratic term of (\ref{S_A-kin}) and the vertex of (\ref{S_A-triple}) with two $V^{(1)}$ and one $A^{(2)}$ contribute to (1b). Similarly (1c) is made of two $\phi^{(2)}$ in (\ref{S_phi-kin}) and two $V^{(1)}$ and $\phi^{(2)}$ in (\ref{S_phi-triple}). The diagram (1d) is the connection of four $V^{(1)}$ via the quad vertex in (\ref{S_V-quad}).

We proceed to the calculation of classical solutions. Let's start with the vector field $V^a_\alpha (Q,z)$. Its equation of motion is
\begin{align*}
 \left[ \p_z \frac{e^{-\tilde{\Phi}}}{z} \p_z g_{\alpha \mu} +  \frac{e^{-\tilde{\Phi}}}{z}(q_1^2 g_{\alpha \mu} - q_{1\alpha} q_{1\mu} ) \right] V_\mu^a (-q_1,z) =
& \frac{e^{-\tilde{\Phi}}}{z} \int d^4 q_2 d^4 q_3 \ i \mathbb{T}^{abc}_{\alpha \beta \gamma} V_\beta^b (q_2,z) V_\gamma^c (q_3,z) .
\end{align*} 
As we are interested in $V^{(1)}$ and $V^{(2)}$ only, we've dropped unnecessary interaction terms. The solution, which is linear in source, is obtained by means of bulk-to-boundary propagator $v(Q,z)$, which is the solution to the linearized equation of motion with unit boundary condition at $z\rar 0$, $v(Q,0)=1$. Using this, we write out the expression for the classical solution:
\begin{gather}
\label{V1}
V^{(1)a}_\alpha (q,z)= \hat{V}^a_\alpha (q) v (q,z).		
\end{gather}
The second order solution can be found perturbatively with the use of the Green function of the equation. Due to the nontrivial vector structure of the equation, we introduce two Green functions, the transverse $G^\perp (Q,z)$  and the longitudinal one $G^\parallel (Q,z)$ with respect to the momentum $Q$. They are solutions to the following equations:
\begin{align}
\label{eq_G}
 \left[ \p_z \frac{e^{-\tilde{\Phi}}}{z} \p_z + \frac{e^{-\tilde{\Phi}}}{z} q_1^2 \right] G^{\perp} (q_1,z,z') &=  \delta(z,z'),\\
\notag
 \p_z \frac{e^{-\tilde{\Phi}}}{z} \p_z  G^{\parallel} (q_1,z,z') &=  \delta(z,z'),
\end{align}
and govern the iteration procedure. For short we will denote their sum by $G^V_{\lambda \lambda'} = \big[g_{\lambda \lambda'} - \frac{Q_\lambda Q_{\lambda'}}{Q^2}\big] G^\perp + \frac{Q_\lambda Q_{\lambda'}}{Q^2} G^\parallel$. Now we can straightforwardly find the second order solution $V^{(2)}$:
\begin{align}
\label{V2}
{V^{(2)}}^a_{\alpha}(-q_1,z) 
=& \int d^4 q_2 d^4 q_3 \ i \mathbb{T}^{abc}_{\alpha' \beta \gamma} \hat{V}_\beta^b (q_2) \hat{V}_\gamma^c (q_3) \times \int dz' \frac{e^{-\tilde{\Phi}}}{z'} G^V_{\alpha \alpha'}(q_1,z,z') v(q_1,z') v(q_2,z').
\end{align}

The equation of motion of dilaton field is
\begin{align*}
 \left[  \p_z \frac{e^{-\tilde{\Phi}-2W}}{z^3} \p_z +  \frac{e^{-\tilde{\Phi}-2W}}{z^3} q_3^2 \right] \ph (-q_3,z)= 
&  \frac{1}{2 \mathcal{A}^2} \frac{e^{-\tilde{\Phi}}}{z} \ \mathbb{D}^{ab}_{\alpha \beta} V_\alpha^a (q_1,z) V_\beta^b (q_2,z).
\end{align*}
As only sources for vector fields are present, we calculate the second order solution immediately. We use the Green function defined by the equation
\begin{equation*}
  \left[ \p_z \frac{e^{-\tilde{\Phi}-2W}}{z^3} \p_z  + \frac{e^{-\tilde{\Phi}-2W}}{z^3} q_3^2 \right] G^{\ph} (q_3,z,z') =  \delta(z,z')
\end{equation*}
and get the solution:
\begin{align}
\label{phi2}
\ph^{(2)} (-q_3,z) &= \int d^4 q_1 d^4 q_2 \ \frac{1}{2 \mathcal{A}^2 }  \int dz'  \frac{e^{-\tilde{\Phi}}}{z'} G^{\ph}(q_3,z,z') \mathbb{D}^{ab}_{\alpha \beta} v(q_1,z') v(q_2,z') \hat{V}_\alpha^a (q_1) \hat{V}_\beta^b (q_2).
\end{align}

The last part is to compute the solution for axial vector field from the equation
\begin{multline}
\label{EOM_A_full}
 \left[ \p_z \frac{e^{-\tilde{\Phi}}}{z} \p_z g_{\gamma \mu} + \frac{e^{-\tilde{\Phi}}}{z}(q_3^2 g_{\gamma \mu} - q_{3\gamma} q_{3\mu}) \right] A_\mu^c(-q_3,z) + \\ 
+ 6 \frac{e^{-\tilde{\Phi}-2W}}{z^3} \chi(z)^2 \Big(A_\gamma^c(-q_3,z) + q_{3 \gamma} \pi(-q_3,z)\Big) = 12 i \ \int d^4 q_1 d^4 q_2 \  \mathbb{A}_{\alpha \beta \gamma}^{abc} V_\alpha^a(z,q_1) V_\beta^b(z,q_2).
\end{multline}
This equation is more complicated than in the vector case, due to the interaction with a scalar profile. We will compute the explicit form of the Green function later. Nevertheless we can use the same notation for its longitudinal and transverse parts as for the vector field and denote it $G^A_{\gamma \gamma'}$. The solution is
\begin{align}
\label{A2}
{A^{(2)}}_\gamma^c (-q_3,z)= \int dz' d^4 q_1 d^4 q_2 \ (12i) \ \mathbb{A}_{\alpha \beta \gamma'}^{abc} \hat{V}_\alpha^a(q_1) \hat{V}_\beta^b(q_2) \ G^A_{\gamma \gamma'}(z,z',q_3)    v(z',q_1) v(z',q_2) 
\end{align}

Having the solutions (\ref{V2}),(\ref{phi2}) and (\ref{A2}) at hand, we are able to write out the classical action of the fourth order in sources. It is
\begin{align}
\notag
S^{(4)} = & \int d^4 q_1 d^4 q_2 d^4 q_3 d^4 q_4 d^4 q_5 \ \delta^4(q_1 {+} q_2 {+} q_5) \ \delta^4(q_5 {-} q_3 {-} q_4) \  \hat{V}_\alpha^a (q_1) \hat{V}_\beta^b (q_2) \hat{V}_\gamma^c (q_3) \hat{V}_\delta^d (q_4)\times \\
\bigg\{ 
\label{VV-V-VV}
  & (-1)\frac{N_c}{24 \pi^2}  \ \mathbb{T}^{abe}_{\alpha \beta \epsilon} \mathbb{T}^{ecd}_{\epsilon' \gamma \delta} \   \times \int dz dz' \frac{e^{-\tilde{\Phi}}}{z} \frac{e^{-\tilde{\Phi}}}{z'}  G^V_{\epsilon \epsilon'} (q_5,z,z') v(q_1,z) v(q_2,z) v(q_3,z') v(q_4,z')  \\
\label{VV-phi-VV}
+ & \frac{1}{4} \frac{N_c}{24 \pi^2} \frac {N_c} {48(N_c^2-1)}   \ \int dz dz' \ \mathbb{D}^{ab}_{\alpha \beta} \mathbb{D}^{cd}_{\gamma \delta} \ \times \frac{e^{-\tilde{\Phi}}}{z} \frac{e^{-\tilde{\Phi}}}{z'} G^{\ph} (q_5,z,z') v(q_1,z) v(q_2,z) v(q_3,z') v(q_4,z') \\ 
\label{VV-A-VV}
- &  6 \frac{N_c}{\pi^2} \int dz dz' \ \mathbb{A}_{\alpha \beta \lambda}^{abe} \mathbb{A}_{\lambda' \gamma \delta}^{ecd}  \times  G^A_{\lambda \lambda'}(z,z',q_3)    v(z',q_1) v(z',q_2) v(z,q_3) v(z,q_4) \\
\label{VVVV}
- &  \frac{1}{2} \frac{N_c}{24 \pi^2} \ \mathbb{Q}^{abcd}_{\alpha \beta \gamma \delta} \ \times \int dz \frac{e^{-\tilde{\Phi}}}{z} v(q_1,z) v(q_2,z) v(q_3,z) v(q_4,z)  \bigg\},
\end{align}
where the following structures appear (we omit here delta-functions, moved to the action):
\begin{align}
\mathbb{T}^{abe}_{\alpha \beta \epsilon} \mathbb{T}^{ecd}_{\epsilon' \gamma \delta} &=  f^{abe} f^{ecd} (g_{\alpha \beta} q_{1\epsilon} + g_{\beta \epsilon} q_{2\alpha} + g_{\alpha \epsilon} q_{5\beta})   (g_{\epsilon' \gamma} q_{5\delta} + g_{\gamma \delta} q_{3\epsilon'} + g_{\epsilon \delta} q_{4\gamma}); \\
\mathbb{D}^{ab}_{\alpha \beta} \mathbb{D}^{cd}_{\gamma \delta} &=   \delta^{ab}  \delta^{cd} \Big\{ \p_{z_1} \p_{z_2} g_{\alpha \beta} +  (q_1 q_2) g_{\alpha \beta} - q_{1\beta} q_{2\alpha}  \Big\} \Big\{ \p_{z_3} \p_{z_4} g_{\gamma \delta} + (q_3 q_4) g_{\gamma \delta} - q_{3\delta} q_{4\gamma}  \Big\}; \\
\label{AA}
\mathbb{A}_{\mu \nu \lambda}^{abe} \mathbb{A}_{\rho \tau \lambda'}^{ecd} &=  \langle t^a t^b t^e \rangle \langle t^c t^d t^e \rangle \epsilon^{\mu \nu \lambda \sigma} \epsilon^{\rho \tau \lambda' \pi} (\p_z^1 q_\sigma^2 - \p_z^2 q_\sigma^1)(\p_z^3 q_\pi^4 - \p_z^4 q_\pi^3);\\
\mathbb{Q}_{\alpha \beta \gamma \delta}^{abcd} &=  f^{abe} f^{ecd} g_{\alpha \gamma} g_{\beta \delta}.
\end{align}
Now taking the variation with respect to four sources resolves into the summation of all 24 permutations of indexes and corresponding momenta in the above structures. Taking the variation with respect to the transverse source will result in the transverse structure in the correlator. We will not write these projectors expicitly and will imply, that the result is transverse with respect to momenta of the corresponding external sources. There is no point if doing the variation in general case, so we will make it for the particular application.

\section{Electric current susceptibility}
\subsection{Kinematics}
One of the interesting problems is the calculation of the two-point function of vector currents in the external magnetic field at vanishing momentum $Q$. In the first order of magnetic field it is related to the correlator of three vector currents, which is zero in our model. To the second order in the magnetic field it can be expressed as
\begin{equation}
\la J_\alpha (Q) J_\beta (-Q)\ra_B = \la J_\alpha(Q) J_\beta(-Q) J_\gamma(k_1) J_\delta(k_2) \ra e_\gamma^1 e_\delta^2|_{k_1,k_2 \rar 0},
\end{equation}
where  $e_\gamma^1, e_\delta^2$ are polarizations of the real photons. We encounter the 4-point correlator, which we can now calculate. As it involves only electromagnetic currents ($J^{em} = \frac{1}{2} J^0 + \frac{1}{6} J^1$) two terms in the variation vanish immediately. These are (\ref{VV-V-VV}) and (\ref{VVVV}), because they contain structure constants $f^{abc}$, which are zero if one of the indices is zero or if any two are identical. Moreover, we see, that the term, that contains intermediate dilaton (\ref{VV-phi-VV}) is suppressed by the factor of $N_c$, comparing to the others. So it is also negligible.

The main input to the correlator is from the axial boson interchange (\ref{VV-A-VV}). Let's take the sum over all permutations of sources $\hat{V}_\alpha^a(Q),\hat{V}_\beta^b(-Q),\hat{V}_\gamma^c(k_1),\hat{V}_\delta^d(k_2)$ in the structure (\ref{AA}). Due to the symmetry properties of this structure all 24 terms form 3 groups of 8, corresponding to S-,T- and U- channels:
\begin{align*}
\mbox{S-channel},\qquad & 2 \langle \{t^a t^b\} t^e \rangle \langle \{t^c t^d\} t^e \rangle \epsilon^{\alpha \beta \lambda \sigma}  \epsilon^{\gamma \delta \lambda' \pi} (\p_z^{Q_1} Q_{2\sigma} - \p_z^{Q_2} Q_{1\sigma})(\p_{z'}^{k_1} k_{2\pi} - \p_{z'}^{k_2} k_{1\pi})\\
\mbox{T-channel},\qquad & 2 \langle \{t^a t^c\} t^e \rangle \langle \{t^b t^d\} t^e \rangle \epsilon^{\alpha \gamma \lambda \sigma}  \epsilon^{\beta \delta \lambda' \pi} (\p_z^{Q_1} k_{1\sigma} - \p_z^{k_1} Q_{1\sigma})(\p_{z'}^{Q_2} k_{2\pi} - \p_{z'}^{k_2} Q_{2\pi})\\
\mbox{U-channel},\qquad & 2 \langle \{t^a t^d\} t^e \rangle \langle \{t^b t^c\} t^e \rangle \epsilon^{\alpha \delta \lambda \sigma}  \epsilon^{\gamma \beta \lambda' \pi} (\p_z^{Q_1} k_{2\sigma} - \p_z^{k_2} Q_{1\sigma})(\p_{z'}^{k_1} Q_{2\pi} - \p_{z'}^{Q_2} k_{1\pi}).
\end{align*}
Here the superscript of the differential $\p_z^Q$ denotes the particular bulk-to-boundary propagator $v(Q,z)$ that should be differentiated. In these expressions we keep only terms linear in $k_1,k_2$ to study the dependence on the magnetic field. Consequently, the S-channel contribution vanishes. We multiply the result by the polarizations of real photons and figure out the dual strength tensor of external field $\epsilon^{\mu \nu \rho \tau} k_\rho e_\tau = \frac{1}{2} \tilde{F}^{\mu \nu}$. Finally, we get 
\begin{align*}
\mbox{T-channel},\qquad &  2 \langle \{t^a t^c\} t^e \rangle \langle \{t^b t^d\} t^e \rangle \left(  \frac{1}{4} \tilde{F}^{\alpha \lambda}(k_1)  \  \tilde{F}^{\beta \lambda'} (k_2) \right) \ \p_z^{Q_1} \p_{z'}^{Q_2},\\
\mbox{U-channel},\qquad &  2 \langle \{t^a t^d\} t^e \rangle \langle \{t^b t^c\} t^e \rangle \left( \frac{1}{4} \tilde{F}^{\alpha \lambda}(k_2)  \  \tilde{F}^{\beta \lambda'} (k_1) \right) \ \p_z^{Q_1} \p_{z'}^{Q_2}. \\
\end{align*}
Our main task is to compute the correlator of electromagnetic currents ($J^{em} = \frac{1}{2} J^0 + \frac{1}{6} J^1$). We can express it via our obtained four-point function:
\begin{align*}
\la J^{em} (Q) J^{em}(Q) J^{em}(k)J^{em}(k) \ra =& \frac{1}{2^4} \la J^{0} J^{0} J^{0}J^{0} \ra \\
& + 2 \frac{1}{2^3} \frac{1}{6} [\la J^{1} J^{0}J^{0}J^{0} \ra + \la J^{0} J^{0} J^{1}J^{0}\ra] \\
& + \frac{1}{2^2} \frac{1}{6^2}[4\la J^{1} J^{0} J^{1}J^{0} \ra + \la J^{1} J^{1} J^{0}J^{0} \ra + \la J^{0}  J^{0} J^{1}J^{1} \ra] \\
& + 2 \frac{1}{2} \frac{1}{6^3} [\la J^{0} J^{1} J^{1}J^{1} \ra + \la J^{1}J^{1}J^{0}J^{1} \ra] \\
& \frac{1}{6^4} \la J^{1}  J^{1} J^{1}J^{1} \ra.
\end{align*}
Having this structure, we can sum up U- and T- channels and get the expression for two-point function of electromagnetic current in the magnetic field.
\begin{align}
\label{JJB}
\la J^{em} (Q) J^{em}(Q)\ra_B =& \left[ \frac{1}{2^4} + 6 \frac{1}{2^2} \frac{1}{6^2} + \frac{1}{6^4} \right] \left( \frac{1}{4} \tilde{F}^{\alpha \lambda}(k_2)  \  \tilde{F}^{\beta \lambda'} (k_1) \right)\\
\notag
& \times (-  6) \frac{N_c}{\pi^2} \int dz dz' \ G^A_{\lambda \lambda'}(z,z',Q)    \p_{z'} v(z',Q) \p_z v(z,Q).v(z,0) v(z',0). 
\end{align}
The number in square brackets comes from the charges calculation and hereafter will be denoted by $\mathcal{C}$.

\subsection{The calculation of the coefficient}

To proceed, we have to specify the model we are working in. The simplest choice is a ``hard-wall'' model proposed in \cite{Erlich,Pomarol}. Other possible choices will be discussed further. In the ``hard-wall'' model there isn't any background dilaton profile or warp factor in the metric, so we put $\tilde{\Phi}$ as zero. The chiral symmetry breaking is described by the scalar profile
\begin{equation*}
\chi(z)=mz+\sigma z^3
\end{equation*} 
in the whole bulk space up to the IR boundary $z_m$. In the limit of $Q \rar 0$ the equations of motion for transverse and longitudinal parts of axial vector field coincide, so the bulk-to-bulk propagator can be expressed as  
\begin{align*}
G^A_{\lambda \lambda'} (Q,z)= g_{\lambda \lambda'}  G(z) + O(Q),
\end{align*}
where $G(z)$ is defined by the equation
\begin{equation*}
\left[\p_z \frac{1}{z} \p_z - 6 \frac{1}{z^3} \chi(z)^2 \right] G(z,z') = \delta(z,z'),
\end{equation*}
with usual hard-wall boundary conditions: 
\begin{equation*}
G(0,z)=0, \qquad \p_z' G(z,z')|_{z'=z_m}=0.
\end{equation*}
The vector bulk-to-boundary propagator can be found explicitly, it's revealed as:
\begin{equation*}
v(Q,z) = \frac{\mathcal{K}_0(Qz_m)}{\mathcal{I}_0(Q z_m)} Q z \ \mathcal{I}_1(Qz) + Q z \mathcal{K}_1(Q z),
\end{equation*}
where $\mathcal{I}$ and $\mathcal{K}$ are modified Bessel functions.
One can check, that at $Q \rar 0$ limit $v(Q,z) = 1$ and $\p_z v(Q,z) \sim Q^2$, so in the integral (\ref{JJB}) only differentials of propagators remain, which result in the overall factor of $Q^4$. Now we can calculate the integral numerically. For the parameters $z_m = (323 Mev)^{-1}$, $m=10 Mev$, $\sigma= (460 Mev)^3$ (see appendix of \cite{Krik2}) we obtain:
\begin{align*}
\la J^{em} (Q) J^{em}(Q)\ra_B \Big|_{HW} =& \mathcal{C} \left( \frac{1}{4} \tilde{F}^{\alpha \lambda}(k_2)  \  \tilde{F}^{\beta}_{\lambda} (k_1) \right) \times (-  6) \frac{N_c}{\pi^2} Q^4 (-1.22 Gev^{-6}). 
\end{align*}
The value of the integral here is computed numerically, it is dimensionful and warrants some more study. In the hard wall model we have 3 dimensionful parameters: $m$,$\sigma$ and $z_m$. One can check, that the quark mass has very little influence on the result, so we can freely take the chiral limit ($m=0$) with the change of about 1\%. The $\sigma$ plays a bigger part, but it is also not
crucial for the result. Taking it equal to zero gives the value $2.93 Gev^{-6}$ and larger $\sigma$ decreases this number until at very large $\sigma \sim 10 GeV^3$ the integral practically vanishes. Important for us is that the current correlator in the external field is nonzero even in the chiral limit and at zero chiral condensate. This indicates, that its origin is not the chiral symmetry breaking. In the case of $m=\sigma=0$ the result obviously behaves as $z_m^6$, or $\Lambda_{QCD}^{-6}$, so taking the scale parameter to infinity kills the result as well. But concerning the limit $\Lambda_{QCD} \rar 0$ we do not find such a smooth behaviour as in the case of condensate. In this limit the integral diverges strongly, which is a sign of qualitative dependence on the $\Lambda_{QCD}$. This lets us conclude, that the most important for the considered value feature of the model is the IR cut-off the AdS space, related to the confinement or conformal symmetry breaking.

According to this issue it is interesting to study the ``soft-wall'' model, because it has an improved behaviour in the IR which includes the proper description of the meson masses \cite{soft,Andreev1}. Moreover, we are not very interested in chiral symmetry braking, which is problematic in considered framework (see \cite{Gherghetta}). In this model one has an exponential factor $e^{cz^2}$ in the action, due to the nontrivial dilaton profile $\Phi$ \cite{soft} or to the deformation of the metric $W$ \cite{Andreev1}. In our case this makes no difference, because it is their sum $\tilde{\Phi}$ which appears in equations of motion. The most popular version of the ``soft-wall'' model has negative $c$ \cite{soft,Gherghetta}, so that the action has a rapidly falling prefactor $e^{-\gamma^2 z^2}$. This gives rise to the strong problem in our calculation, because the bulk-to-bulk propagator, being the inverse of the kinetic term in the action, acquires a factor $e^{\gamma^2 z^2}$ and diverges extremely fast at large $z$. This makes the diagram under consideration (\ref{VV-A-VV}) diverge too, because the vertices coming from the Chern-Simons term do not have any exponential suppression, due to the independence of metric and dilaton. As a consequence, one can not use the perturbation theory here. This problem is not surprising, since the model with negative $c$ does not express the confinement (namely the proper heavy quark potential) crucial for our computation \cite{Andreev2,Brodsky}. 

In order to incorporate the confinement, one has to use the model with positive $c$ \cite{Andreev2,Brodsky,Fen}. Then all the solutions to equations of motion decay in IR as well as the bulk-to-bulk propagator. So the perturbation calculation converges and we can get the striking result:
\begin{align*}
\la J^{em} (Q) J^{em}(Q)\ra_B \Big|_{SW} =& \mathcal{C} \left( \frac{1}{4} \tilde{F}^{\alpha \lambda}(k_2)  \  \tilde{F}^{\beta}_{\lambda} (k_1) \right) \times (-  6) \frac{N_c}{\pi^2} (0.7 Gev)^{-2}. 
\end{align*}
It is nonzero at $Q=0$, what seems to be quite unphysical in the framework of zero temperature and absence of any chemical potentials. The reason is the presence of the unphysical massless mode in the vector sector in the model with positive $c$, which has been observed in \cite{soft} and discussed in \cite{Fen}. This feature prevents us from trusting the answer above, until the nature of this mode is clarified.

\section{Conclusion}
In this work we've performed a perturbative 5D calculation of the vector current four-point function of QCD via the holographic model. The generality of the used setting allows us to study some common features of such calculations in various AdS/QCD models. Namely we find, that due to the normalisation of dilaton field, the exchange of dilaton in the bulk is suppressed by the factor of $N_c$. Even though the form of the dilaton field and its features in holography are still subjects of investigation, we claim, that this behaviour has to remain in all the models. The obtained result is applicable to a large class of commonly used holographic models and can be used in various problems, as it includes the chiral symmetry breaking term, the anomaly term, the arbitrary number of quark flavors, the arbitrary dilaton and scalar potentials and the arbitrary warp factor in the metric.

We use the vector current four-point function to calculate the two-point correlator of electromagnetic currents in the external magnetic field. This quantity turns out to be weakly dependent on the chiral symmetry breaking in the model and is sensitive to its infrared behaviour, namely to the way the conformal symmetry is broken. Studying this calculation in the hard wall AdS/QCD model and soft wall models with different dilaton behaviour, we find that the perturbation theory is divergent in the soft wall model with falling exponent in the action. This problem is not visible until every term of the action, including interactions, is multiplied by the same factor. But considering the Chern-Simons action, which is free of metric as well as dilaton input, we get the exponentially growing bulk-to-bulk propagator, not suppressed by the interaction vertices and leading to the strong divergence of the result. This turns to a handicap for calculating diagrams with bulk-to-bulk propagators and Chern-Simons vertices in such a framework. In the model with growing exponent in the action we obtain the value, finite in the limit of small $Q$, which is unphysical at zero temperature. This results from the presence of the massless pole in the vector two-point function, which leads to the violation of charge conservation and is forbidden in QCD. We hope, that the nature of this pole and the solution of this defect will allow more accurate calculations in this model.

In the end of the day we obtain the best result in the framework of the ``hard-wall'' model. The value of the correlator in the magnetic field is proportional to $B^2$ at small $B$, where $B$ is the field strength.  At small $Q$ it goes to zero as $Q^4$, what is consistent with general expectations at zero temperature and chemical potentials. This confirms that the holographic model, featuring chiral symmetry breaking and chiral anomaly, does not autonomously develop the chiral chemical potential, needed for the nonzero result. Some other components are obviously necessary to describe the phenomena of heavy ion collisions. Constructing a holographic model which would feature nonzero temperature and topological charge of QCD could shed light on this problem.

\section{acknowledgments}
I am grateful to Alexander Gorsky for suggesting this problem and general guidance. I would also like to thank Peter Kopnin for useful discussions. This work was partially supported by Russian President's Grant for Support of Scientific Schools NSh-3036.2008.2, by RFBR grant 09-02-00308 and Dynasty Foundation. 

\end{document}